# Quantum memory in warm rubidium vapor with buffer gas

Mark Bashkansky*, Fredrik K. Fatemi, and Igor Vurgaftman

*Naval Research Laboratory, Optical Sciences Division, 4555 Overlook Avenue S.W., Washington, D.C. 20375*
*Corresponding author: bashkansky@nrl.navy.mil*

The realization of quantum memory using warm atomic vapor cells is appealing because of their commercial availability and the perceived reduction in experimental complexity. In spite of the ambiguous results reported in the literature, we demonstrate that quantum memory can be implemented in a single cell with buffer gas using the geometry where the write and read beams are nearly co-propagating. The emitted Stokes and anti-Stokes photons display cross-correlation values greater than 2, characteristic of quantum states, for delay times up to 4 μs. © 2011 Optical Society of America

The implementation of quantum memory (QM) via the interaction between atomic ensembles and optical fields attracted considerable attention ever since the DLCZ (Duan, Lukin, Cirac, and Zoller) protocol was proposed in 2001 [1]. The original DLCZ scheme is based on the correlation between the emission of a Stokes photon and the collective excitation in an atomic ensemble produced by a spontaneous Raman scattering event. Since the original DLCZ proposal, QM in atomic ensembles has been demonstrated in both cold and warm vapors [2,3], the latter being particularly appealing because of commercial availability and the perceived reduction in experimental complexity. However, previous reports employ widely different experimental conditions and geometries and occasionally suffer from ambiguous results. Some of the work is focused on the storage of single photons produced elsewhere rather than on the generation of correlation in a single cell [4-8].

The divergence of experimental conditions in the reports claiming quantum memory can be illustrated as follows [4,5,9-11]. In one work, a 4-mm diameter write (read) beam with an intensity of $I \approx 10^{-4}$ ($10^{-3}$) W/m$^2$ was utilized in a room-temperature $^{87}$Rb vapor cell (atomic density $N = 1.3 \times 10^{10}$ cm$^{-3}$) with 30 Torr Ne buffer gas [9]. Although quantum correlations between Stokes and anti-Stokes photons were originally reported, a later Erratum withdrew this claim [9]. A study by a different group [10] employed $^{87}$Rb vapor at 75°C ($N = 1.1 \times 10^{12}$ cm$^{-3}$) with 3 Torr Ne buffer gas and a 100 μm-wide write beam with intensities in the range of $I = 3 \times 10^3 - 1.3 \times 10^4$ W/m$^2$, a two-order of magnitude difference in the number density and a six-order of magnitude difference in the write-beam intensity by comparison with [9]. As discussed in [10], the experiment was not performed in the single-photon regime required for the DLCZ protocol.

While the presence of a buffer gas is necessary to reduce atomic diffusion and enable a sufficiently long-lasting QM, another study [11] pointed out the presence of collisionally redistributed fluorescence (CRF) caused by buffer-gas collisions. In that work, with $^{87}$Rb vapor at 60°C ($N = 3.3 \times 10^{11}$ cm$^{-3}$), 7 Torr Ne buffer gas, and write (read) beam $I = 1.4(7) \times 10^4$ W/m$^2$, the fluorescence was noted to limit the fidelity of QM severely. The maximum observed cross-correlation between Stokes and anti-Stokes photons was 1.3 (still classical and no specified time delay), which implies that the noise was too high for the implementation of the DLCZ protocol.

Whereas the two studies discussed above [9,10] employed co-propagating write and read beams, the counter-propagating geometry has also been reported in the literature[4,5] and in fact claimed to be optimal for the observation of correlations [12]. In that work, Stokes and anti-Stokes photons were generated in the source cell without any reported time delay, and the latter were stored and regenerated at a later time in a second (target) cell using electromagnetically induced transparency. However, the storage of the collective atomic excitation in the source cell was not explored.

In this work we present what is to the best of our knowledge the first unambiguous demonstration of quantum memory in a single warm vapor cell and address the salient issues in the literature discussed above. We operate in the nearly single-photon regime desired for QM by reducing the Raman excitation probability. Spectral filtering is improved, and both the write and read beams are detuned by larger amounts than in the previous works. We demonstrate that quantum correlations between Stokes and anti-Stokes photons can be maintained for a few microseconds, which has not been shown before in a single atomic cell.

Figure 1(a) illustrates our experimental setup, which is similar to previous works [9-11]. The 7.5-cm-long $^{87}$Rb cell is held at 37°C ($N = 4.3 \times 10^{10}$ cm$^{-3}$), with 1 or 10 Torr of the Ne buffer gas. The D$_1$ ($5^2S_{1/2} \rightarrow 5^2P_{1/2}$) line of $^{87}$Rb used in the present experiment is schematically shown in Fig. 1(b). Our write (read) beams, with powers of 0.6 (1.2) mW, were collimated with a waist of 1.3 mm, which corresponds to the relatively low $I \approx 1.1(2.2) \times 10^2$ W/m$^2$. Write pulses of 1 μs duration were detuned 1.3 GHz below the F'= 1 level, and the higher-frequency write beam was filtered by a heated $^{85}$Rb cell in a magnetic field, without any noticeable effect on the co-propagating Stokes photons. Read pulses of 1 μs duration and variable delay after the write pulses were detuned 1.08 GHz above

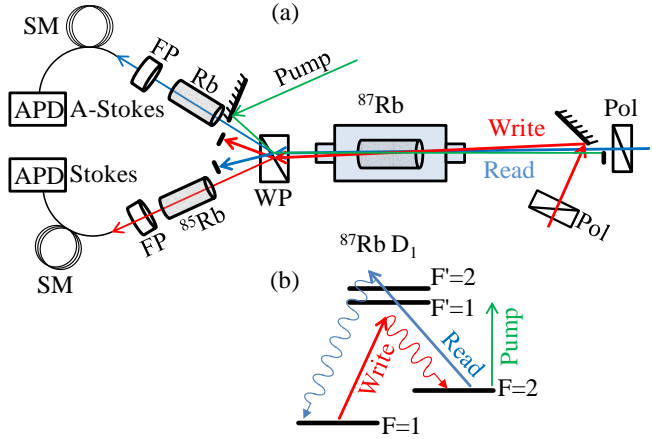

Fig. 1. (Color online) (a) Simplified experimental setup showing magnetically shielded $^{87}$Rb vapor cell, held at 37°C and containing Ne buffer gas either at 1 Torr or 10 Torr. Pol – Glan Thompson polarizer, WP – Wallaston prism, FP – Fabry Perot filters, APD – avalanche photodiode, SM – single mode fiber. (b) Simplified 87 Rubidium D1 level diagram.

the F ' = 2 level and filtered out by another heated Rb cell.

The pump pulses, tuned to the F=2 → F'=1 transition, started a few microseconds after the read pulses and ended 400 ns before the write pulses. The repetition rate of 20 kHz was employed for both spectral scans and correlation measurements. The write and read beams intersected at $\theta_{WR}$ = 6 mrad, and Fabry-Perot filters with finesse of ≈ 100 and transmission bandwidths of 100 MHz and 130 MHz were used for Stokes and anti-Stokes photons, respectively. After the $^{87}$Rb cell, the overall transmission efficiencies to the Perkin Elmer SPCM-AQR-14 avalanche photodiodes (APD) were ≈ 30% for Stokes and ≈15% for anti-Stokes photons and the APD efficiency, dark count rate and dead-time were 60%, 100 cps, and 80 ns, respectively. The nearly single-photon regime is realized by making the experimental conditions much more stringent than those of [11], e.g. only ≈ 0.005 Stokes photons per shot are detected. An attempt to perform the experiment under the conditions specified in [9] did not produce an observable signal.

The presence of CRF [11] in a cell with 10 Torr of Ne was confirmed for Stokes and anti-Stokes beams separately by introducing relatively large $\theta_{WS}$ = $\theta_{RAS}$ = 6 mrad. Figure 2 shows the spectral scans of the Stokes and anti-Stokes channels, with the pump beam absent for the latter. The write (read) beams were detuned $\delta_W(\delta_R)$ = 1.2(1.085) GHz below (above) the resonance. CRF is observed on both sides of the weaker signal owing to the finite free spectral range of the etalons, but when the pump is present both the anti-Stokes signal and fluorescence are strongly reduced. The scans in Fig. 2 show clearly that for the large detunings used, the signal is well resolved from the fluorescence. The best signal-to-noise ratio (defined as the ratio of the Stokes to the fluorescence counts) of 10 is achieved for $\delta_W$ = 1.3 GHz and degrades rapidly for $\delta_W$ < 1 GHz.

The scans for the cells with 1 and 10 Torr Ne buffer gas and otherwise identical conditions are compared in Fig. 3. The fluorescence to the F = 2 level is suppressed by the absorption in the magnetically broadened and heated $^{85}$Rb filter cell. In contrast to the previous work [11], even though the fluorescence signal is much lower in the 1 Torr Ne cell, the Stokes signal is increased only marginally. Since the present geometry supports many spatial modes ($\approx$ 8000), a large change in the amount of CRF emitted into those modes need not correspond to a similar and opposite change in the Stokes signal emitted into the forward direction.

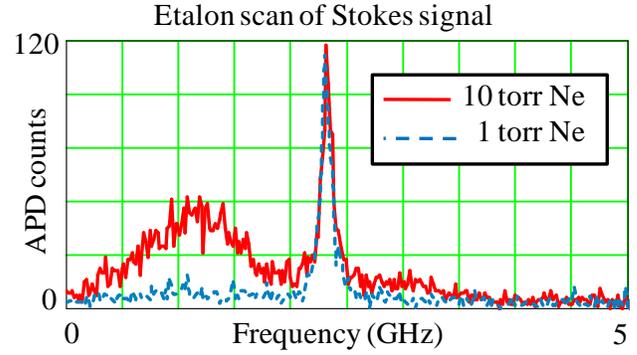

Fig. 3. (Color online) Spectral scan of the Stokes signal for 10 Torr (Solid, red) and 1 Torr (dashed, blue) Ne cells. The scans take 1 second per point, the cell temperature is 37°C, the write pulse duration is 1 µs, and the detuning is 1.3 GHz.

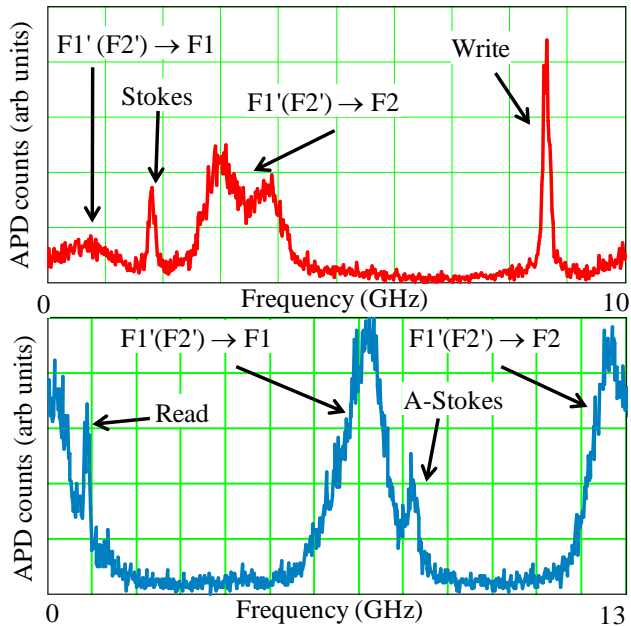

Fig. 2. (Color online) Spectral scans using tunable etalons of the Stokes (above) and anti-Stokes (below) channels. The cell temperature was 37°C, the write and read pulse durations were 2 µs and the detunings 1.2 and 1.085 GHz, respectively.

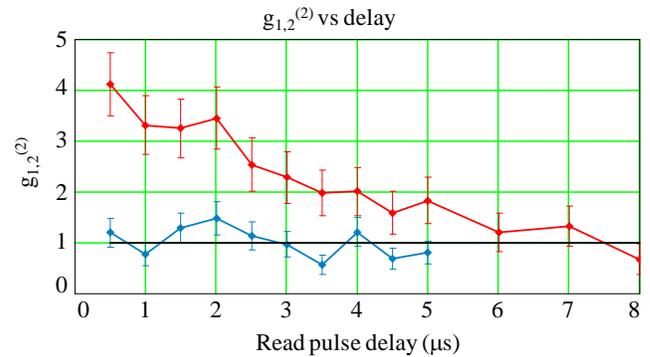

Fig. 4. (Color online) Measured intensity correlation between Stokes and anti-Stokes photons as a function of the delay between the write and read 1 µs long pulses for the 10 Torr Ne cell. The upper curve is for the etalon set to transmit anti-Stokes photons, while the lower curve is for the etalon tuned to transmit the collisional anti-Stokes fluorescence. The error bars denote the calculated 1 σ deviation values.

The cross-correlation between the Stokes and anti-Stokes photons $g_{1,2}^{(2)} = <:n_1n_2:> / <n_1><n_2>$, where normal operator ordering is indicated, is necessary to establish the quantum nature of the generated fields. Figure 4 shows the measured $g_{1,2}^{(2)}$ as a function of delay time between the write and read pulses. Each point represents 500 s of data collection. We estimate that only $p_s = 0.5\%$ of Stokes photons, $p_s^2 = 2.5 \times 10^{-5}$ of the Stokes two-photon emission, and 0.02% of all (correlated and uncorrelated) anti-Stokes photons per shot are detected.

Since by design we operate in the nearly single-photon regime, the lower cell temperature, larger detunings, and larger beam sizes, required for longer QM, result in reduced efficiency and longer collection time. However, these limitations are not fundamental and can be overcome by using higher optical powers and improved filtering. Correlations $g_{1,2}^{(2)} > 2$, obtained up to $\approx 4$ $\mu$s storage, are nearly always indicative of non-classical correlations between Stokes and anti-Stokes fields [3], although a formal demonstration would require verifying that $g_{1,1}^{(2)}, g_{2,2}^{(2)} \leq 2$. The cross correlation $g_{1,2}^{(2)} > 1$ is maintained for $\approx 8$ $\mu$s.

To determine whether the CRF during the read process is correlated with Stokes photons, we detuned the anti-Stokes etalon to transmit the fluorescence instead of the anti-Stokes signal (lower curve in Fig. 4). Since no cross-correlation is observed, the fluorescence can be removed by spectral filtering without affecting the fidelity of QM. It likely arises from the other spatial modes stored in the atomic ensemble during the writing or from the four-wave-mixing [5] undergoing collisional redistribution during the read process.

The loss of Stokes photons to fluorescence simply reduces the duty cycle, since only detected photons are assumed to correspond to stored atomic excitations. As shown here, fluorescence can be overcome with larger detunings. It was also demonstrated that larger detunings have other benefits such as the storage of shorter pulses [7, 8]. For on-resonance anti-Stokes photons, the contribution of the fluorescence photons to the signal can directly affect QM. If the read beam is tuned off resonance, and the four-wave-mixing is suppressed [5], the impact of the fluorescence should also be minimal.

Geometric considerations play an important role in the design of the experiments. The storage time after writing depends critically on the angle between the write beam and Stokes photons $\theta_{WS}$ for large-size beams in a long warm vapor cell. The effective wavelength of the atomic spin wave is given by $\Lambda \approx 795$ nm/sin($\theta_{WS}$) so that even $\theta_{WS} = 1$ mrad leads to $\Lambda < 1$ mm, which is still usable for compact cold-atom traps, but unacceptable for warm vapors even in the presence of buffer gas.

Furthermore, for any beams wider than a few hundred microns, the angle between the read and the anti-Stokes beams should be close to zero as well. While a relatively small angle between the write and read beams does not impact QM negatively so long as the beams overlap over the full interaction region in the cell, the counter-propagating geometry does not satisfy the phase-matching condition for anti-Stokes generation $\boldsymbol{k}_S + \boldsymbol{k}_{AS} = \boldsymbol{k}_W + \boldsymbol{k}_R$ [3] and appears suitable only for cold-atom traps.

In summary, we have unambiguously demonstrated non-classical correlations in a single warm vapor cell, and clarified under what conditions the writing, storage, and retrieval necessary for the implementation of QM can be demonstrated in cells containing buffer gas. Additionally, we have addressed several points of significant confusion in the literature. We have further shown that the presence of collisional fluorescence in cells with buffer gas need not be detrimental.

This work is supported by the Office of Naval Research.